\documentclass[twocolumn, letter]{aastex631}
\usepackage{amssymb}
\usepackage{xcolor}
\usepackage{colortbl}
\usepackage{hyperref} %% to avoid \citeads line fills
\newcommand{\sfr}{$\rm M_{\odot}~yr^{-1}$}
\newcommand{\ergs}{erg s$^{-1}$}
\newcommand{\msun}{M$_{\odot}$}
\newcommand{\lsun}{L$_{\odot}$}
\newcommand{\kms}{km s$^{-1}$}

\newcommand{\lprime}{K \kms pc$^{2}$}

\newcommand\aastex{AAS\TeX}
\newcommand\latex{La\TeX}

\shorttitle{Molecular gas in the z=7.5 quasar Pōniuā‘ena}
\shortauthors{Feruglio, C. et al.}

\graphicspath{{./}{figures/}}

\begin{document}
	
	\title{First constraints of dense molecular gas at z$\sim$7.5 from the quasar Pōniuā‘ena}
 
	\author[0000-0002-4227-6035]{Chiara Feruglio}
	\correspondingauthor{C. Feruglio}
	\email{chiara.feruglio@inaf.it}
	\affiliation{INAF - Osservatorio Astronomico di Trieste, Via G. Tiepolo 11, I-34143 Trieste, Italy}
	\affiliation{IFPU - Institute for Fundamental Physics of the Universe, via Beirut 2, I-34151 Trieste, Italy}

    \author[0000-0002-0039-3102]{Umberto Maio}
	\affiliation{INAF - Osservatorio Astronomico di Trieste, Via G. Tiepolo 11, I-34143 Trieste, Italy}

	\author[0000-0002-9909-3491]{Roberta Tripodi}
	\affiliation{Dipartimento di Fisica, Università di Trieste, Sezione di Astronomia, Via G.B. Tiepolo 11, I-34131 Trieste, Italy}
	\affiliation{INAF - Osservatorio Astronomico di Trieste, Via G. Tiepolo 11, I-34143 Trieste, Italy}
	\affiliation{IFPU - Institute for Fundamental Physics of the Universe, via Beirut 2, I-34151 Trieste, Italy}

	\author[0000-0001-6114-9173]{Jan Martin Winters}
	\affiliation{IRAM, 300 rue de la Piscine, Domaine Universitaire de Grenoble, 38406 St.-Martin-d’Hères,France}

	\author[0000-0002-4205-6884]{Luca Zappacosta}
	\affiliation{INAF - Osservatorio Astronomico di Roma, Via Frascati 33, I-00040 Monte Porzio Catone, Italy}
 
	\author[0000-0002-4314-021X]{Manuela Bischetti}
	\affiliation{Dipartimento di Fisica, Università di Trieste, Sezione di Astronomia, Via G.B. Tiepolo 11, I-34131 Trieste, Italy}
	\affiliation{INAF - Osservatorio Astronomico di Trieste, Via G. Tiepolo 11, I-34143 Trieste, Italy}

\author[0000-0002-2115-1137]{Francesca Civano}
\affiliation{Center for Astrophysics | Harvard \& Smithsonian, Cambridge, MA 02138}
 
	\author[0000-0002-6719-380X]{Stefano Carniani}
	\affiliation{Scuola Normale Superiore, Piazza dei Cavalieri 7 I-56126 Pisa, Italy}
	
	\author[0000-0003-3693-3091]{Valentina D'Odorico}
	\affiliation{INAF - Osservatorio Astronomico di Trieste, Via G. Tiepolo 11, I-34143 Trieste, Italy}
	\affiliation{IFPU - Institute for Fundamental Physics of the Universe, via Beirut 2, I-34151 Trieste, Italy}
	\affiliation{Scuola Normale Superiore, Piazza dei Cavalieri 7 I-56126 Pisa, Italy}
	
	\author[0000-0002-4031-4157]{Fabrizio Fiore}
	\affiliation{INAF - Osservatorio Astronomico di Trieste, Via G. Tiepolo 11, I-34143 Trieste, Italy}
	\affiliation{IFPU - Institute for Fundamental Physics of the Universe, via Beirut 2, I-34151 Trieste, Italy}
 
	\author[0000-0002-7200-8293]{Simona Gallerani}
	\affiliation{Scuola Normale Superiore, Piazza dei Cavalieri 7 I-56126 Pisa, Italy}
	
	\author[0000-0002-9122-1700]{Michele Ginolfi}
	\affiliation{Dipartimento di Fisica e Astronomia, Universit'a di Firenze, Via G. Sansone 1, 50019, Sesto Fiorentino (Firenze), Italy }
	
	\author[0000-0002-4985-3819]{Roberto Maiolino}
	\affiliation{Institute of Astronomy, University of Cambridge, Madingley Road, Cambridge CB3 0HA, UK}
	\affiliation{Kavli Institute for Cosmology, University of Cambridge, Madingley Road, Cambridge CB3 0HA, UK}
	\affiliation{Department of Physics and Astronomy, University College London, Gower Street, London WC1E 6BT, UK}
	
	\author[0000-0001-9095-2782]{Enrico Piconcelli}
	\affiliation{INAF - Osservatorio Astronomico di Roma, Via Frascati 33, I-00040 Monte Porzio Catone, Italy}
	
	\author[0000-0003-3050-1765]{Rosa Valiante}
	\affiliation{INAF - Osservatorio Astronomico di Roma, Via Frascati 33, I-00040 Monte Porzio Catone, Italy}

 	\author[0000-0001-7883-496X]{Maria Vittoria Zanchettin}
	\affiliation{SISSA, Via Bonomea 265, 34136 Trieste, Italy}
 	\affiliation{INAF - Osservatorio Astronomico di Trieste, Via G. Tiepolo 11, I-34143 Trieste, Italy}
	
	\begin{abstract}
		We report the detection of CO(6-5) and CO(7-6) and their underlying continua from the host galaxy of quasar J100758.264+211529.207 (Pōniuā‘ena) at z=7.5419, obtained with the NOrthern Extended Millimeter Array (NOEMA). 
  Pōniuā‘ena belongs to the HYPerluminous quasars at the Epoch of ReionizatION (HYPERION) sample of 17 $z>6$ quasars selected to be powered by supermassive black holes (SMBH) which experienced the fastest mass growth in the first Gyr of the Universe.
   The one reported here is the highest-redshift measurement of the cold and dense molecular gas to date. The host galaxy is unresolved and the line luminosity implies a molecular reservoir of $\rm M(H_2)=(2.2\pm0.2)\times 10^{10}$ \msun, assuming a CO spectral line energy distribution typical of high-redshift quasars and a conversion factor $\alpha=0.8$ $\rm M_{\odot} (K\,km \, s^{-1} \,pc^{2})^{-1} $. We model the cold dust spectral energy distribution (SED) to derive a dust mass of M$_{\rm dust} =(2.1\pm 0.7)\times 10^8$ \msun, and thus a gas to dust ratio $\sim100$. Both the gas and dust mass are not dissimilar from the reservoir found for luminous quasars at $z\sim6$. 
  We use the CO detection to derive an estimate of the cosmic mass density of $\rm H_2$, $\Omega_{H_2} \simeq 1.31 \times 10^{-5}$. This value is in line with the general trend suggested by literature estimates at $ z < 7 $ and agrees fairly well with the latest theoretical expectations of non-equilibrium molecular-chemistry cosmological simulations of cold gas at early times.

	\end{abstract}
	
	%% Keywords should appear after the \end{abstract} command. 
%% The AAS Journals now uses Unified Astronomy Thesaurus concepts:
%% https://astrothesaurus.org
%% You will be asked to selected these concepts during the submission process
%% but this old "keyword" functionality is maintained in case authors want
%% to include these concepts in their preprints.

%% From the front matter, we move on to the body of the paper.
%% Sections are demarcated by \section and \subsection, respectively.
%% Observe the use of the LaTeX \label
%% command after the \subsection to give a symbolic KEY to the
%% subsection for cross-referencing in a \ref command.
%% You can use LaTeX's \ref and \label commands to keep track of
%% cross-references to sections, equations, tables, and figures.
%% That way, if you change the order of any elements, LaTeX will
%% automatically renumber them.
%%
%% We recommend that authors also use the natbib \citep
%% and \citet commands to identify citations.  The citations are
%% tied to the reference list via symbolic KEYs. The KEY corresponds
%% to the KEY in the \bibitem in the reference list below. 

\section{Introduction} \label{sec:intro}

Since the first $z>6$ quasar discovery \citep{fan2001}, the population of quasars near the Epoch of Reionisation (EoR) has increased up to $\sim 300$ sources and the frontier of the quasar search has been pushed back to 0.7 Gyr with the recent discovery of eight $z>7$ quasars. Noteworthy, three of them are at $z\sim7.5$, well inside the EoR \citep{Banados2018Nat,yang2020,wang2021,fan2022}. 
These quasars are powered by SMBH with masses from $10^8$ $\rm M_\odot$ up to $10^{10}$ $\rm M_\odot$, shining close to the Eddington limit, with bolometric luminosities at the brightest end of the quasar luminosity function, $\rm L_{bol}>5 \times 10^{46}$ erg s$^{-1}$ \citep{willott2010,mazzucchelli2017,onoue2019}. 
At Far-IR/sub-mm wavelengths, observations reveal the presence of copious amounts of dust ($> 10^8$ M$_\odot$), and star formation rates (SFR) up to 1000-3000 M$_\odot$ yr$^{-1}$, within the host galaxies \citep[e.g.]{maiolino2005,wang2013,feruglio2018, venemans2017a,venemans2020}, albeit with large systematic uncertainties \citep{Tripodi2023}.

The cold molecular phase is the least explored up to now at $z>6$. A few tens of quasars at redshift up to $z\sim6.8$ have been detected in carbon monoxide (CO) rotational transitions, that are the most direct tracers of the cold molecular interstellar medium (ISM),indicating massive molecular reservoirs of dense gas feeding both star formation and nuclear accretion \citep{wang2013,wang2016,gallerani2014,carniani2019,venemans2017a, decarli2022}, with a combination of photo-dissociation (PDR) and X-ray dominated regions (XDR) \citep{li2020,Pensabene2021,decarli2022}).
Only in a few cases it was possible to spatially resolve the molecular reservoirs and map disks or dispersion-dominated hosts \citep{walter2004,feruglio2018,Tripodi2022}. 

Currently only eight quasars are known at redshift $z>7$ of which three at $z\sim7.5$ \citep[and refs. therein]{fan2022}. In two of these quasars, J1342+0928 at $z=7.54$ and J112001.48+064124.30 at $z=7.08$, the dense molecular reservoirs have been explored and remain undetected \citep{novak2019}.

This Letter is the second of a series of papers (the first being \citet{Tripodi2023} dedicated to the HYPerluminous quasars the Epoch of ReionizatION (HYPERION) sample. HYPERION consists of seventeen $z=6-7.5$ luminous ($\rm L_{bol}\sim 10^{47.3}$ \ergs) quasars selected to be powered by SMBH which experienced the fastest mass growth in the first Gyr of the Universe, and that are targets of a 2.4 Ms XMM-Newton Multi-Year Heritage programme (Zappacosta et al. in prep.).
In this Letter we report the detection of CO(6-5) and (7-6) along with their underlying continua for the quasar J100758.264+211529.207 (dubbed Pōniuā‘ena) at $z=7.5419$, obtained with the NOrthern Extended Millimeter Array (NOEMA) in the framework of a multiwavelength follow-up of the HYPERION quasars. 
%This Letter is the second in a series of papers \citep[e.g.]{Tripodi2023} centered on the sub-mm properties of the HYPERION quasar sample, which includes the 18 QSOs that experienced the most rapid SMBH mass growth at $z>6$ discovered so far.
This quasar was first discovered by \citet{yang2020} and is one of the three highest redshift quasars known, all located at the midpoint of the Reionisation Epoch, $z\sim7.5$. 
The CO detections presented here enable the first estimate of the molecular gas reservoirs at these high redshifts, and of the gas to dust mass ratio. 
We adopt a $\Lambda$CDM cosmology with parameters $\Omega_\Lambda=0.714$, $\Omega_m=0.286$ and $H_0=69.6$. Thus, the angular scale is $\sim 5.11$ kpc/arcsec at $z\sim 7.5$.

\section{Observations and Data Analysis} \label{sec:obs}

Observations were taken with the NOEMA interferometer under project W21ED in March 2022. The receivers were tuned at 81.2 GHz in the Lower Side Band (LSB). 
The CO(6-5) emission line from the quasar host is redshifted to the LSB, while the CO(7-6) line lies in the Upper Side Band (USB), so the bandwidth covered by the correlator Polifyx enables  detection of both lines in one frequency tuning.
Amplitude/phase calibrators are J1012+232 and J0953+254, and LkHa101 (0.2 Jy) was used as flux calibrator. 
Calibration and imaging was performed using CLIC and MAPPING within the GILDAS software (www.iram.fr/IRAMFR/GILDAS).

%The system introduces parasite current at 94744.44  MHz (USB) which produces a strong narrow signal, close to the frequency expected for the CO(7-6) emission line from the quasar. We exclude therefore the channel corresponding to 94744.44  MHz from the following analysis. 

%\section{Analysis} \label{sec:ana}

The continuum visibility tables at two representative frequencies in LSB and USB were derived using the task \text{uv\_filter} within MAPPING to filter out a spectral region 400 MHz wide around the emission lines, and \text{uv\_continuum}. The noise reaches 15.5 $\mu$Jy/beam in LSB and 13.7 $\mu$Jy/beam in USB over a bandwidth of 7.4 GHz in each sideband (excluding the 400 MHz spectral window conataining the emission lines).
The continuum uv-tables were analyzed in the uv plane. We find flux densities of $57\pm 14$ $\mu$Jy at 81.2 GHz and $87\pm14$ $\mu$Jy at 94 GHz, with both measurements consistent with point sources (Tab. \ref{tab1}).
Deconvolution with the Hogbom cleaning algorithm without applying any mask produced a clean beam of $\rm 3.1\times 2.2~ arcsec^2$ (PA=30 deg) in LSB and $\rm 2.5\times 1.8~ arcsec^2$ (PA=30 deg) in USB.

\begin{figure*}[t]
	\centering
	\includegraphics[width=0.9\linewidth]{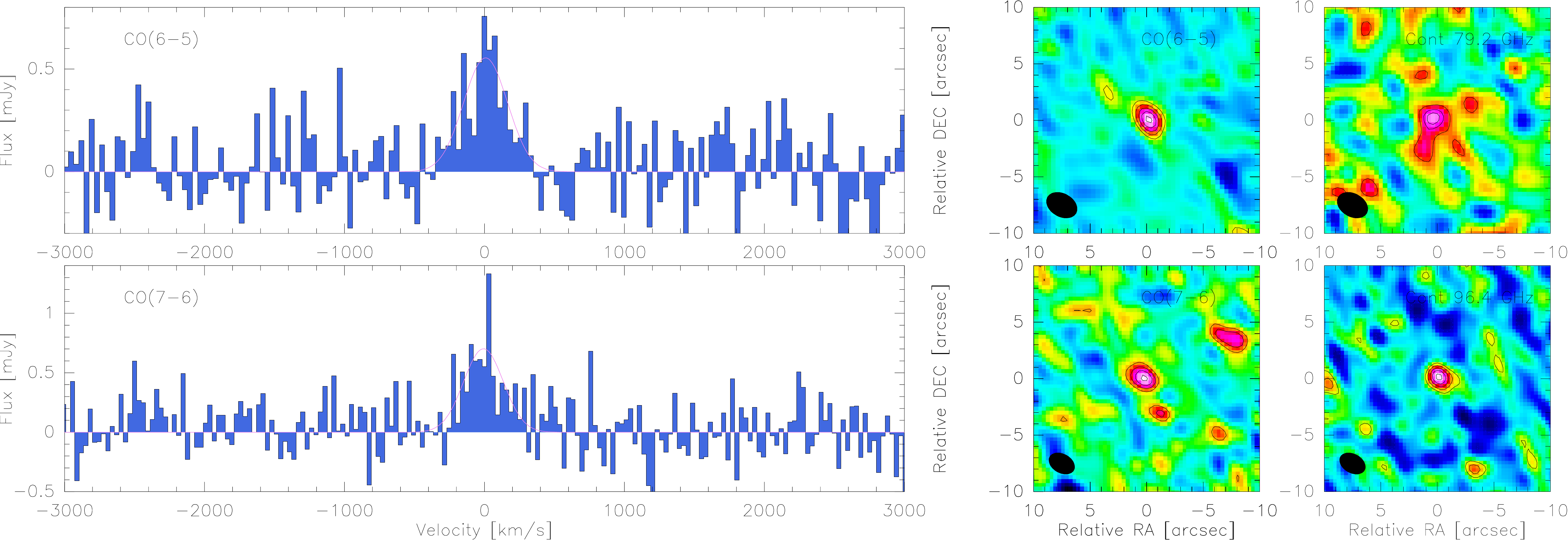}
	\caption{\footnotesize Upper panels: (left to right) the CO(6-5) emission line, the corresponding velocity-integrated CO map and the  79.2 GHz continuum map of Pōniuā‘ena. Lower panels: the CO(7-6) emission line, the corresponding velocity-integrated map and the 96.4 GHz continuum map. Magenta lines show the fit with a single Gaussian component, FWHM is reported in Table \ref{tab1}. Contours are drawn starting at 2$\sigma$ in steps of 1$\sigma$, $\sigma=0.06$ Jy km s$^{-1}$ for CO(6-5) and 0.05 Jy km s$^{-1}$ for CO(7-6). }
	\label{fig:lines}
\end{figure*}

The CO(6-5) and (7-6) line profiles and velocity-integrated maps were produced by subtracting the continuum in each sideband using \textit{uv\_subtract} (Fig. \ref{fig:lines}).  Noise levels are 0.28 mJy/beam per 74 \kms channel in LSB and 0.26 mJy/beam per 63 \kms channel in USB. 
We produced an averaged uv table across the line width for both lines, using the \textit{uv\_average} task within MAPPING, and analyzed the line visibilities. 
Both lines are consistent with an unresolved source in the uv plane with fluxes $Sdv_{\rm CO6-5}=0.44\pm0.06$ Jy \kms and $Sdv_{\rm CO7-6}=0.40\pm0.07$ Jy \kms (Tab. \ref{tab1}). 
Both lines are robustly detected with statistical significance of 7 and 6 $\sigma$ respectively.
A fit with a Gaussian model gives a marginally resolved source and a lower signal-to-noise ratio in the flux, confirming that the source is unresolved in both transitions. 

The average redshift derived from CO(6-5) and (7-6) lines is z$_{\rm CO}=7.5149\pm0.0006$, consistent with that derived from [CII]$\lambda$158 $\mu$m \citep{yang2020}. 
The line widths, derived by fitting a single Gaussian to the spectra, are FWHM$_{\rm CO(6-5)}=371\pm61$ \kms and FWHM$_{\rm CO(7-6)}=334\pm56$ \kms, consistent with the [CII] FWHM \citep{yang2020}. 
The line luminosities are $\rm L^\prime CO(6-5)=(2.1\pm0.3)\times 10^{10}$ \lprime and $\rm L^\prime CO(7-6)=(1.38\pm0.24)\times 10^{10}$ \lprime\ \citep{carilli2013}.  
The [CI](2-1) 370$\mu$m emission line is undetected and we derive a 3$\sigma$ upper limit on the flux of 0.18 Jy \kms, or L[CI]$<10^8$ \lsun (Table \ref{tab1}), assuming an unresolved source with FWHM=350 \kms. 
A scan of the data cubes did not reveal any other line or continuum emitters.

\begin{table}[b]
\vspace{0.2cm}
		\caption{Properties of Pōniuā‘ena}
		\centering
		\begin{tabular}{lc}
			\hline
			\hline
RA & 10:07:58.279 \\
DEC & +21:15:28.932 \\
z$_{\rm [CII]}^{(a)}$& $7.5149\pm0.0004$  \\
M$_{\rm BH}^{(a)}$ [ \msun] & $(1.5\pm0.2)\times 10^9$ \\ 
\hline 
F$_{\rm cont,79.2GHZ}$ [$\mu$Jy]& $57\pm14$  \\
F$_{\rm cont,96.4GHZ}$  [$\mu$Jy] & $87\pm14$\\
\hline
z$_{\rm CO}^{(b)}$ &  $7.5149\pm0.0006$\\
FWHM$_{\rm CO(6-5)}$ [\kms] & $371\pm61$  \\ 
F$_{\rm CO(6-5)}$ [Jy \kms] &  $0.44\pm 0.06$  \\
L$^\prime_{\rm CO(6-5)}$ [ \lprime]&  $(2.1\pm 0.3)\times 10^{10}$  \\
L$_{\rm CO(6-5)}$ [L$_\odot$] &  $(2.2\pm 0.3)\times 10^{8}$  \\
FWHM$_{\rm CO(7-6)}$ [\kms] &  $334\pm56$  \\
F$_{\rm CO(7-6)}$ [ Jy \kms] & $0.40\pm 0.07$ \\
L$^\prime_{\rm CO(7-6)}$ [\lprime] &  $(1.38\pm 0.24)\times 10^{10}$  \\
L$_{\rm CO(7-6)}$ [L$_\odot$]&  $(2.3\pm 0.4)\times 10^{8}$ \\
F$_{\rm [CI]}$ [\rm Jy~ \kms]&  $<0.18$ \\
L$^\prime$[CI] [\lprime]& $<6\times10^9$ \\  
L[CI] [L$_\odot$]&  $<1.0\times 10^8$ \\
\hline
M(H$_2)^{(c)}$ [\msun]& $(2.2\pm0.2)\times 10^{10}$ \\
M$_{\rm dust}$ [\msun] &  $(2.1\pm 0.7)\times 10^8$ \\
GDR$^{(d)}$ & 105 \\
  $\beta$ & $1.53\pm0.17$ \\
  SFR$^{(e)}$ [\sfr]&  165\\
\hline
		\end{tabular}
		\label{tab1}
		\flushleft 
		\footnotesize {{\bf Notes.} All fluxes are derived from fit of the visibilities with a point source model. (a) \citet{yang2020}. (b) Average of CO 6-5 and 7-6 redshifts. (c) Assuming a conversion factor $\alpha_{\rm CO}$ = 0.8 $\rm M_{\odot} (K\,km \, s^{-1} \,pc^{2})^{-1} $} and assuming a brightness temperature ratio $\rm r_{61}=CO(1-0)/CO(6-5)=0.75$ (d) Gas to dust ratio.
  (e) The SFR is computed assuming a dust temperature $T_{\rm dust}=50$ K (see Sec. \ref{sec:disc}) and is corrected by a factor of 50\%, taking into account the contribution of the QSO to the dust heating \citep{duras2017}. 
  	\end{table}

\begin{figure}[b]
	\centering
	\includegraphics[width=\linewidth]{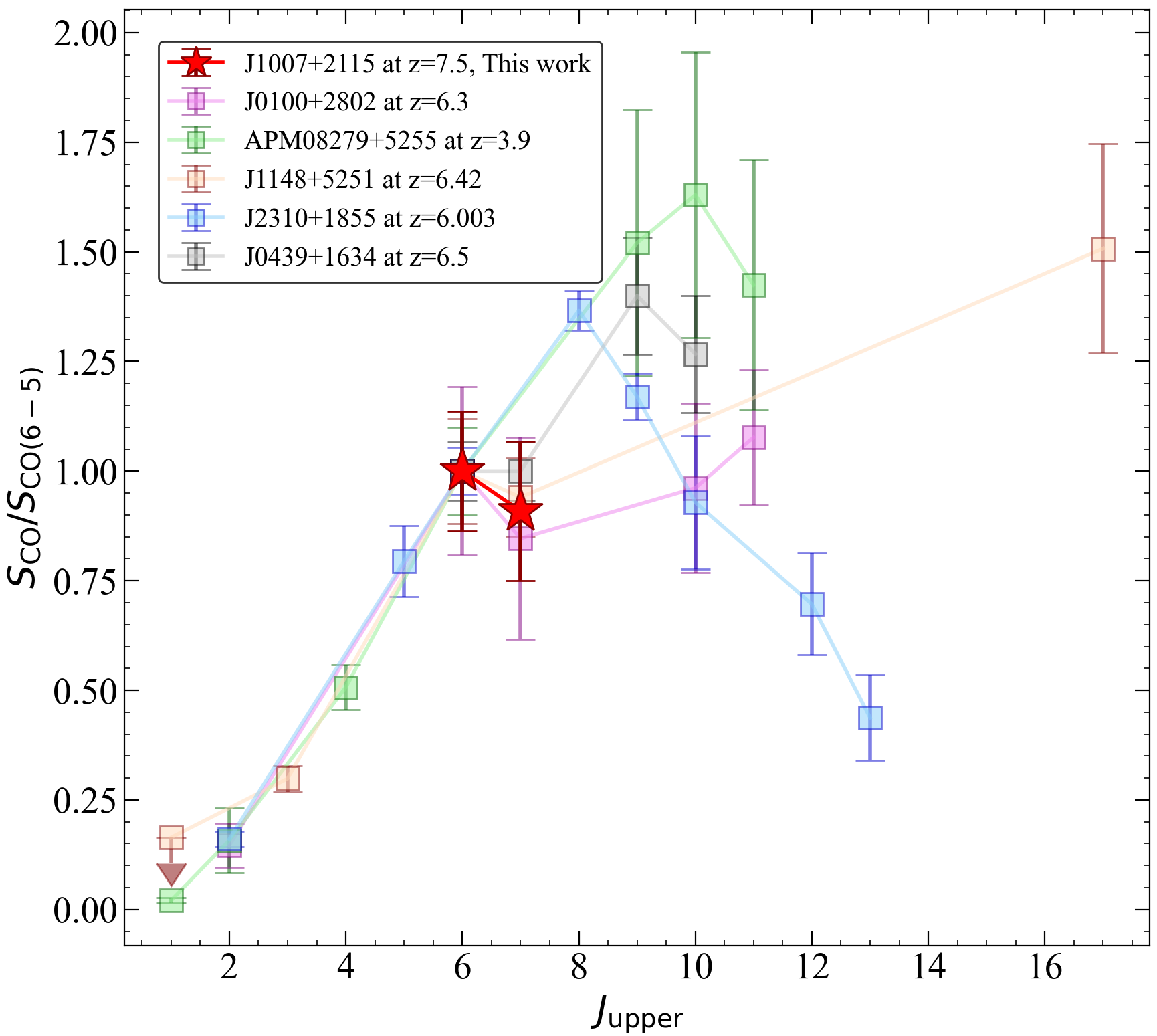}
	\caption{\footnotesize CO SLED of Pōniuā‘ena compared with those of other QSOs at lower redshift. The CO SLED for Pōniuā‘ena is shown as red stars, for J0439+1634 at $z=6.5$ as grey squares \citep{Yang2019}, for J1148+5251 at $z=6.42$ as orange squares \citep{riechers2009, gallerani2014}, for J0100+2802 at $z=6.3$ as purple squares \citep{wang2019}, for J2310+1855 at $z=6.003$ as blue squares \citep{li2020} and for APM08279+5255 at $z=3.9$ as green squares \citep{papado2001, weiss2007}.}
	\label{fig:sled}
\end{figure}

\begin{figure*}[t]
	\centering
	\includegraphics[width=0.95\linewidth]{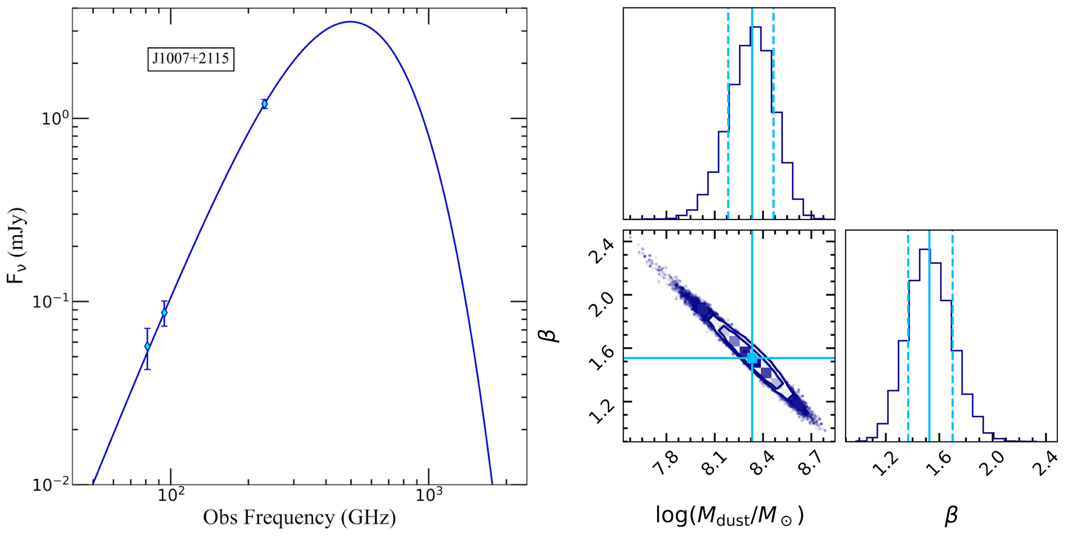}
	\caption{\footnotesize Results of the SED fitting of Pōniuā‘ena. Left panel: SED using our new NOEMA data at 3mm (i.e., $\sim 80$ and $\sim 90$ GHz), and the ALMA data at $\sim 230$ GHz taken from \citet{yang2020}. The best-fitting curve with dust temperature fixed at $50$ K is shown as a blue solid line. Right panel: Corner plot showing the two dimensional posterior probability distributions of $M_{\rm dust}, \beta$. Cyan solid lines indicate the best-fitting parameter, while the dashed lines mark the 16\% and 84\% percentiles for each parameter.}
	\label{fig:sed}
\end{figure*}

\section{Discussion and Conclusions} \label{sec:disc}

We use the CO line detections to provide the first measurement of cold molecular gas mass reservoirs at redshift $\sim7.5$, in the host galaxy of quasar Pōniuā‘ena.  
Pōniuā‘ena belongs to the HYPerluminous quasars at the Epoch of ReionizatION (HYPERION) sample of $z>6$ QSOs, that have been selected to include the quasars with the most massive SMBH at their epoch, possibly resulting from exceptionally fast mass growth during their accretion history. The HYPERION sample consists of 17 quasars with $L_{\rm bol} = 10^{47.3}$ erg/s, SMBH masses in the range $M_{\rm BH} = 10^{9}-10^{10}\ \rm M_\odot$, and Eddington ratios $>0.3-0.4$. The HYPERION survey will be presented in Zappacosta et al. (in prep.).

The CO Spectral line energy distribution (SLED) of Pōniuā‘ena, limited to the (6-5) and (7-6) transitions, is shown in Fig. \ref{fig:sled}. We show Pōniuā‘ena together with other $z>6$ quasars and quasar APM08279+5255 at $z=3.9$, for which the SLED is well constrained from J=2 up to J=17 \citep{li2020,wang2019,riechers2009,gallerani2014,weiss2007}. 
The SLEDs have been normalised to the J=6-5 transition. 
For Pōniuā‘ena the SLED shows a flattening at the CO(6-5) and (7-6) transitions,  similarly to what is observed in other QSOs at $z\sim6$, J0100+28, J1148+52 and J0439+1643 \citep{wang2019, gallerani2014,Yang2019,carniani2019}. For those quasars, the CO SLED suggests two gas components with $T_{\rm kin}\sim 20$ K and $\sim200 $ K, of which the coldest one is associated to the dust heated by star formation and the warmest is likely heated by the AGN. Additional rotational CO transitions would be needed to assess whether this is the case also for Pōniuā‘ena. 

Based on the typical CO SLED of $z\gtrsim6$ quasars (Fig. \ref{fig:sled}), we adopt a r61=CO(6-5)/CO(1-0)= 0.75 luminosity ratio, and a conversion factor $\alpha_{\rm CO}=0.8 \rm M_{\odot}~ (K\,km\,s^{-1} \,pc^{2})^{-1}$ \citep{carilli2013}, to derive the molecular gas mass from the CO(6-5) line luminosity. With these assumptions we find a molecular mass $\rm M(H_2)=\alpha_{\rm CO}\times L^\prime~ r61^{-1}=(2.2\pm 0.2)\times 10^{10}$ \msun. If instead we used the average SLED of $z\sim2.5$ galaxies in ASPECS \citep[r61=0.28,][]{boogard2020}, this would imply a $\times 3$ larger molecular mass.
%$M(H_2)=\sim 6\times 10^{10}$ \msun. 
On the other hand, if we used the L$^\prime$CO(7-6) and r71=0.17, as done for the sample of quasars at z=5-6.7 in \citet{decarli2022}, we would obtain a $\times 3$ lower $M(H_2)$. 
The typical SLED of quasars is steeper than that of galaxies \citep{li2020}, but the CO SLED is the main systematic uncertainty at play.  

Applying the relation by \citet{zanella2018} to convert [CII] luminosity \citep{yang2020} into molecular mass would yield $\rm M(H_{2,[CII]})\sim 5.7\times 10^{10}$ \msun.
Our fiducial value, $\rm M(H_2)=(2.2\pm 0.2)\times 10^{10}$ \msun, is at the lowest end of the range found for $z>6$ quasars, and the luminosity ratio $\rm L[CII]/LCO(7-6)=7$ for Pōniuā‘ena while the mean value is around 20 for $z>6$ quasars \citep{feruglio2018,wang2019,decarli2022}. 

We estimate the total dynamical mass of the system by $\rm M_{\rm dyn}\approx M(H_2)+M_{HI}+M_{\rm BH}$, where $\rm M_{HI}$ is the atomic gas mass associated with PDR and derived from [CII], $\rm M_{HI}=1.44\times 10^9$ \msun \citep{Hailey2010}, $\rm M_{\rm BH}$ is the black hole mass from \citet{yang2020}, and we neglect the stellar component. We find $\rm M_{\rm dyn}\sim 2.7\times 10^{10}~\rm M_\odot$, where the molecular mass fraction is $\rm M(H2)/M_{\rm dyn}\sim 80\%$. Should the stellar component be significantly massive, as discussed in \citet{valiante2014}, the gas fraction would decrease to $\lesssim 40\%$. A more accurate estimate of the dynamical mass would require [CII] or CO observations that spatially resolve the host galaxy.

\begin{figure*}[t]
	\centering
	\includegraphics[width=0.75\linewidth]{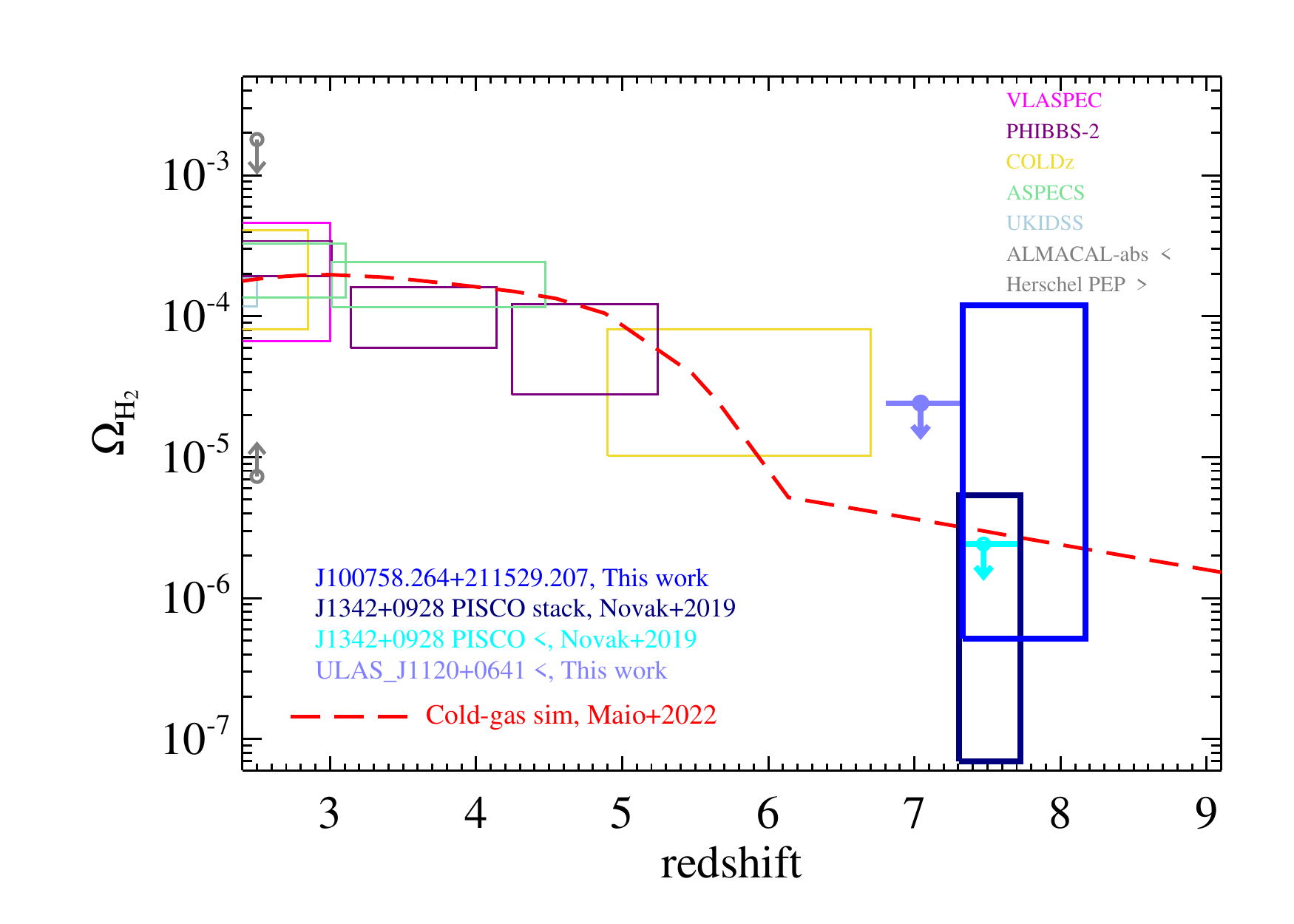}
	\caption{\footnotesize Comoving cosmic mass density of cold molecular gas as a function of redshift. Vertical sizes indicate the uncertainties in each bin. Data are from
\citet{riechers2020a,riechers2020b} (VLASPEC, COLDz), \citet{decarli2020} (ASPECS), \citet{lenkic2020} (PHIBSS-2), \citet{garratt2021} (UKIDSS-UDS), and \citet{Hamanowicz2023} (ALMACAL-CO), and upper/lower limits from ALMACAL-abs \citep{klitsch2019} and Herschel PEP \citep{berta2013}. Theoretical expectation from cold-gas simulations (dashed line) are by \citet{maio2022}.
}
	\label{fig:omegah2}
\end{figure*}

The Spectral Energy Distribution (SED) of the cold dust component of Pōniuā‘ena, based on these observations and ALMA Band 6 observations from \citet{yang2020}, is shown in Figure \ref{fig:sed}. 
We model the dust continuum with
a modified black-body (MBB) function given by
 \begin{equation}\label{eqsed}
     S_{\nu_{\rm obs}}^{\rm obs} = \frac{\Omega}{(1+z)^3}[B_{\nu}(T_{\rm dust}(z))-B_{\nu}(T_{\rm CMB}(z))](1-e^{-\tau_{\nu}})
 \end{equation}

\noindent where $\Omega = (1+z)^4A_{\rm gal}D_{\rm L}^{-2}$ is the solid angle with $A_{\rm gal}$ the surface area and and $D_{\rm L}$ the luminosity distance of the galaxy, respectively. The dust optical depth is given by 
    $\tau_{\nu}=\frac{M_{\rm dust}}{A_{\rm galaxy}}k_0 (\frac{\nu}{250\ \rm GHz})^{\beta}$,
 with $\beta$ the emissivity index and $k_0 = 0.45\  \rm cm^{2}\ g^{-1}$ the mass absorption coefficient \citep{beelen+2006}. 
 The effect of the CMB on the dust temperature is taken into account as $B_{\nu}(T_{\rm CMB}(z)=T_0(1+z))$ \citep{dacunha2013}. Since the source is unresolved, the adopted area of the galaxy is the size of the synthetic beam in the Band 6 observation \citep{yang2020}, $0.4\times 0.3$ arcsec, and it is equivalent to considering the area of a circle with 1 kpc radius. 

Dust temperature cannot be constrained with the low-frequency data in hand, hence we fix it to a value found on average in quasars at this redshift, $T_{\rm dust}=50$ K \citep{Tripodi2023, Tripodi2022, carniani2019,wang2019}. We explore the two dimensional parameter space using a Markov chain Monte Carlo (MCMC) algorithm implemented in the \texttt{EMCEE} package \citep{foreman2013}, assuming uniform priors for $M_{\rm dust}$ and $\beta$. We derive a cold dust mass of $M_{\rm dust}=2.1\pm 0.7\times 10^8\ \rm M_{\odot}$ and a dust emissivity index of $\beta=1.53 \pm 0.17$. Varying the temperature up to $T_{\rm dust}=80$ K, the dust mass becomes a factor of two lower, and the emissivity index slightly decreases to $\beta = 1.27$. 

The gas-to-dust ratio is about GDR $\sim 105$, in line with QSOs at lower redshift \citep{Tripodi2022,bischetti2021}.  
We compute the SFR using the relation by \citet{kennicutt1998} scaled to a Chabrier IMF: SFR$(M_{\odot}/{\rm yr})=10^{-10}\ L^{\rm IR}_{\ 8-1000 \mu m}(L_\odot)$. We also take into account the contribution of the luminous QSO to the dust heating with a factor of 50\% \citep{duras2017}, and we obtain SFR=165 \sfr, in agreement with the broad range suggested by \citet{yang2020}. However this value has a large systematic uncertainty since the dust temperature is not determined. 
At face value this would imply a star formation efficiency of $\rm SFE=SFR/M_{H_2}=7.5~ 10^{-9}$ yr$^{-1}$.
Observations in ALMA Band 9 are needed to possibly resolve the host galaxy and tightly constrain both $T_{\rm dust}$ and SFR \citep{Tripodi2023}. 

Finally, we derive an estimate of the cosmic mass density of molecular mass, $\Omega_{\rm H_2}$, at z$\sim 7.5$.
Thanks to the wide band covered by our observation, we probe the redshift 
range between 7.33 and 8.17, or the range 600-700 Myr of cosmic time,  
the field of view is taken as the primary beam of the observation, about 50 arcsec.
We use the estimated molecular mass $\rm M(H_2)$ in the corresponding cosmic volume, $V$, to derive   
the H$_2$ mass density parameter 
$
 \Omega_{\rm H_2} = \rm M(H_2) / V / \rho_{\rm crit, 0} 
$, 
where 
$ \rho_{\rm crit, 0} \simeq 277.4 \, h^2\, \rm  M_{\odot } / kpc^3 $ 
is the present-day cosmological critical density and $h = H_0 / (100~{\rm km/s/Mpc})$.
 
Fig.~\ref{fig:omegah2} shows the $\Omega_{\rm H_2}$ redshift evolution derived from this observation and those of two other quasars at $z>7$ for which CO upper limits are measured, i.e. J1342+0928 \cite[for which we also report the tentative stack detection by][]{novak2019}, and ULAS J1120+0641 (ALMA archive).
Literature values at lower $z$ are from VLASPEC, COLDz 
\citep{riechers2020a,riechers2020b}, ASPECS \citep{decarli2020}, PHIBSS-2 \citep{lenkic2020}, UKIDSS-UDS \citep{garratt2021}, and ALMACAL-CO \citep{Hamanowicz2023} , and upper (lower) limits from ALMACAL-abs \citep{klitsch2019} and Herschel PEP \citep{berta2013}.
The value inferred by our analysis of Pōniuā‘ena is 
$ \Omega_{\rm H_2} \simeq 1.31 \times 10^{-5} $.
%
% 1.3144560e-05 in [1.2188592e-05, 1.4100528e-05] statistical range
%
Upper and lower limits are evaluated by considering a statistical error on H$_2$ mass determination of $ 0.16 \times 10^{10} \,\rm M_\odot $, a systematic calibration error of 10\%, and CO sled lower and upper errors of 0.65 and 5.94$\times 10^{10} \,\rm M_\odot$.
For J1342+0928 \citep{novak2019} the resulting upper limit between $z \simeq 7.30$ and 7.72 
% $z \simeq 7.305$ and 7.725 
suggests $ \Omega_{\rm H_2} < 2.42 \times 10^{-6} $, 
% 2.4194000e-06
while the stacking analysis gives 
$ \Omega_{\rm H_2} \simeq 1.13 \times 10^{-6}$.
%  1.1364744e-06
In this latter case, a statistical error on H$_2$ mass of $2.2 \times 10^8 \,\rm M_\odot $ and an upper limit of a factor of 2 for the stacking error have been considered. 
The ALMA archive observation of ULAS J1120+0641 allows us to estimate an $ \Omega_{\rm H_2}$ upper limit of $ 2.42 \times 10^{-5}$ 
%   2.4161000e-05
at $ z\simeq 6.80 - 7.13$.
% $ z\simeq 6.805 - 7.135$.

As a comparison, we also show the trend expected by the latest, accurate, non-equilibrium molecular-chemistry cosmological simulations of cold gas at early times by \cite{maio2022}.
The predicted $ \Omega_{\rm H_2} $ behaviour at $ z > 6 $ is mainly driven by H$_2$ formation via H$^-$ channel, since in the simulation the dust growth is inefficient at such primordial epochs.
At later times the effects of UV radiation (that enhances production of free charges at temperatures around or below $ 10^4 \,\rm K$) and dust grain catalysis in progressively enriched media boost $ \Omega_{\rm H_2}$ expectations.
Overall, the values we find are in line with the general trend suggested by literature estimates at $z<7$ and agree fairly well with the latest theoretical expectations.
Although $ \Omega_{\rm H_2} $ determinations by quasar data might be slightly biased, as individual objects do not necessarily represent a fair sample of the Universe, our results suggest that it is possible to leverage on this by combining different objects at similar cosmological epochs.
We note that this work represents the first attempt to set constraints on H$_2$ abundances by combining state-of-the-art interferometric observations of the cold dense molecular gas in the first 700 Myr with state-of-the art cold-gas modelling.
Previous works have indeed either neglected a fully complete modeling of primordial molecules or could not rely on constraints from observational data for the early regimes probed here.

\begin{acknowledgments}
\textit{Acknowledgements.} This publication has received funding from the European Union’s Horizon 2020 research and innovation programme under grant agreement No 101004719 (ORP). Authors acknowledge financial support from PRIN MIUR contract 2017PH3WAT; PRIN MAIN STREAM INAF "Black hole winds and the baryon cycle"; the Horizon 2020 INFRAIA Programme under Grant Agreement n. 871158 AHEAD2020; INAF Large Grant 2022 "Toward an holistic view of the Titans: multi-band observations of $z>6$ QSOs powered by greedy supermassive black-holes". This work is co-funded by the European Union (ERC, WINGS, 101040227). Views and opinions expressed are however those of the author(s) only and do not necessarily reflect those of the European Union or the European Research Council Executive Agency. Neither the European Union nor the granting authority can be held responsible for them. This work is based on observations carried out under project number W21ED with the IRAM NOEMA Interferometer. IRAM is supported by INSU/CNRS (France), MPG (Germany) and IGN (Spain).
This paper makes use of the following ALMA data: ADS/JAO.ALMA\#2019.1.01025.S. ALMA is a partnership of ESO (representing its member states), NFS (USA) and NINS (Japan), together with NRC (Canada), MOST and ASIAA (Taiwan) and KASI (Republic of Korea), in cooperation with the Republic of Chile. The Joint ALMA Observatory is operated by ESO, AUI/NRAO and NAOJ.
\end{acknowledgments}

\bibliography{biblio}{}
\bibliographystyle{aasjournal}

\end{document}